\documentclass[a4paper]{article}
\usepackage{tabularx}
\usepackage{amsmath,graphicx}
\usepackage{multicol}
\usepackage{multirow}
\usepackage{INTERSPEECH2021}

\title{BeamTransformer: Microphone Array-based Overlapping Speech Detection}
\name{Siqi Zheng, Shiliang Zhang, Weilong Huang, Qian Chen, Hongbin Suo, Ming Lei, Jinwei Feng, Zhijie Yan}
\address{Speech Lab, Alibaba Group}
\email{\{zsq174630, sly.zsl, yuankai.hwl\}@alibaba-inc.com}

\begin{document}

\maketitle
\begin{abstract}
  We propose BeamTransformer, an efficient architecture to leverage beamformer's edge in spatial filtering and transformer's capability in context sequence modeling. BeamTransformer seeks to optimize modeling of sequential relationship among signals from different spatial direction. Overlapping speech detection is one of the tasks where such optimization is favorable. In this paper we effectively apply BeamTransformer to detect overlapping segments. Comparing to single-channel approach, BeamTransformer exceeds in learning to identify the relationship among different beam sequences and hence able to make predictions not only from the acoustic signals but also the localization of the source. The results indicate that a successful incorporation of microphone array signals can lead to remarkable gains. Moreover, BeamTransformer takes one step further, as speech from overlapped speakers have been internally separated into different beams. 
  
\end{abstract}
\noindent\textbf{Index Terms}: Overlapping Speech Detection, Speaker Diarization, Microphone Array, Beamforming

\section{Introduction}

Overlapping speech detection (OSD) is a key component for speaker diarization and speech separation. Speaker diarization seeks to match a time frame of speech to the corresponding speaker identity. The agglomerative hierarchical clustering on speaker embeddings has been one of the main approaches for speaker diarization \cite{DBLP:journals/taslp/MiroBEFFV12}\cite{DBLP:conf/icassp/Garcia-RomeroSS17}\cite{DBLP:conf/icassp/WangDWMM18}\cite{DBLP:journals/corr/abs-2012-14952}. Audios are split into segments and speaker embeddings are extracted from each of the segments. Unsupervised clustering is performed on the speaker embeddings to obtain speaker identities\cite{DBLP:conf/interspeech/ZhengLSL19a}. One problem of clustering approach is that speaker embeddings extracted from small segments can be biased by the speech content \cite{DBLP:conf/interspeech/ZhengLS20}. Furthermore, it lacks the ability to handle overlapping speech.

Recently an end-to-end speaker diarization approach is proposed, which incorporates overlapping speech detection into the process of inferring speaker labels at any given frame time \cite{DBLP:conf/asru/FujitaKHXNW19}\cite{DBLP:conf/interspeech/FujitaKHNW19}\cite{DBLP:conf/interspeech/HoriguchiF0XN20}\cite{DBLP:journals/corr/abs-2003-02966}. When tested on datasets with overlapping speech, the end-to-end approach achieved a better performance than the embedding-based clustering methods. The improvement is not unexpected since the baseline clustering method does not have the capability to handle overlapping speech, which makes up a considerable proportion of the test set. It remains unclear how accurate the system is in detecting overlapping speech. Given its importance in the speaker diarization system, it is worthwhile to single out the OSD component for extensive study.

When overlapping speech occur, identifying speakers of interest is not enough. It is also necessary to separate the overlapping speech in order to obtain the clean automatic speech recognition (ASR) results corresponding to the target speakers. Therefore, a speech separation or target speaker extraction component is usually applied to overlapping speech segments. 

A vast number of literatures can be found on  speech separation \cite{DBLP:conf/interspeech/WangLSWCLHLPNG20}\cite{DBLP:conf/icassp/DelcroixZKON18}\cite{DBLP:journals/taslp/LuoM19}\cite{DBLP:journals/taslp/WangNW14}\cite{DBLP:journals/taslp/WangC18a}. A major challenge faced by researchers studying speech separation is that it may cause distortions to original speech and result in degradation to the performance of ASR \cite{DBLP:conf/interspeech/WangMWSWHSWJL19}. This poses as a discouraging factor for real application since overlapping speech only contributes to a small portion of the entire duration and harming the rest of the non-overlapping speech may seem unworthy. Furthermore, performing multi-target speaker extraction on non-overlapping speech may result in undesired ``ghost speech", which will contaminate the final ASR deliverable. Therefore, knowing when to perform speech separation is crucial. 

In this paper we aim to detect overlapping speech segments on microphone array data. A microphone array-based beamforming process separates speech into multiple beams according to the localization of the corresponding speakers. Therefore, when an overlapping segment is detected, separated speech is produced alongside. The only further action required is to select the optimal beam(s), depending on most probable localization of the speaker(s). \cite{anguera2007acoustic} is one of first works focusing on beamforming approach for speaker diarization. It provided a non-neural network based approach to take advantage of multi-microphone data. In \cite{DBLP:conf/interspeech/ZhengS21} we proposed a speaker diarization system based on microphone array data. We showed that integrating spatial spectrum information can lead to remarkable improvement to the system. An OSD component was mentioned in the proposed system but was not elaborated. In this paper we discuss in details the OSD component mentioned in previous work. We propose a neural network architecture called BeamTransformer, which manages to take advantage of beamformer's ability to utilize multi-microphone data and transformer's power in context sequence modeling.

Several previous efforts on detecting overlapping speech detection is worthy to be addressed. \cite{DBLP:conf/interspeech/YoshiokaECXA18} proposed a BLSTM based unmixing transducer that transforms an input multi-channel acoustic signal into a fixed number of time-synchronous audio
streams. The AMI corpus is one of the few publicly available corpus containing annotations to generate labels for overlapping speech detection \cite{mccowan2005ami}. \cite{DBLP:conf/interspeech/AndreiCB17} is among one of first efforts to train deep neural networks on OSD task. The authors reported a 0.77 precision and 0.68 recall on 25ms frame on AMI corpus. \cite{DBLP:conf/specom/KunesovaHZR19} seeks to improve speaker diarization by detecting overlapping speech and reports a 0.71 precision and 0.48 recall. \cite{DBLP:conf/icassp/BullockBG20} claims to have made significant improvements, observing a 0.92 precision and 0.48 recall on AMI. 

According to the performance reported by the aforementioned works, frame-level overlapping speech detection is still far from being a reliable component in the system of speaker diarization and automatic speech recognition. It is suspected that a 25ms frame of input feature does not contain enough useful information for the neural network to make reliable prediction on the OSD task. One frame of information may be sufficient for a ``0 vs 1" task, such as voice activity detection, but the ``1 vs more" task requires more.

We choose to experiment on one-second segments instead of frames for several reasons. First, inferring on small segments allow us to take into account the sequence context information, which turns out to be helpful for inputs from multiple beams. The segment-level classification reports a more convincing result than the frame-level performance mentioned above. With accuracy, recall, and precision hovering around 90 percent, we can be more confident on relying on the outputs of OSD component in practical speaker diarization and ASR systems. 

Second, the final goal of overlap detection is not simply inferring speaker labels, but to separate and recognize speech from the targeted speakers of concern. Overlapping speech are usually processed by target speaker extraction and speech separation, both of which require segment-level inputs. Hence a frame-level responsiveness is trivial. 

Third, frame-level OSD task requires accurate labeling on every frame, which is highly impractical. \cite{DBLP:conf/interspeech/AndreiCB17} reported that AMI corpus contains labelling errors in terms of OSD annotations. Segment-level OSD task, on the other hand, is much more tolerant on the quality of annotations.

\section{Methods}

This section is organized as follows. In 2.1 we discuss the methods used to extract acoustic and spatial features from microphone array data. In 2.2 the architecture of BeamTransformer is thoroughly discussed. In 2.3 a complementary component used to process spatial features is described. 

\subsection{Features}
The acoustic and spatial features we utilize are extracted from our previously proposed Circular Differential Directional Microphone Array (CDDMA). The detailed design of CDDMA are more thoroughly discussed in \cite{DBLP:conf/interspeech/ZhengS21}\cite{huangdifferential}. The look-direction of each beamformer is uniformly distributed around a circle to cover the whole space. The output signals of the beamformers are spatially separated from each other. All the directional elements are uniformly distributed on a circle and directions are pointing outward. The CDDMA beamformer is given as below:
\begin{equation}\label{solution}
	\textbf{h}_{cddma}(\omega)  = \textbf{R}^{H}( \omega, \boldsymbol{\theta }) [\textbf{R}( \omega, \boldsymbol{\theta }) \textbf{R}^{H}( \omega, \boldsymbol{\theta })]^{-1} \textbf{c}_{\boldsymbol{\theta }} . 
\end{equation}
where the vector $\textbf{c}_{\boldsymbol{\theta }}$ defines the acoustic properties of the beamformer such as beampattern; the constraint matrix $\textbf{R}( \omega, \boldsymbol{\theta })$ of size $\mathit{N} \times \mathit{M} $ is constructed by the directional microphone steering vector which exploits the acoustics of microphone elements. As shown in \cite{huangdifferential}, the CDDMA-beamformer displays significant improvement in terms of white noise gain (WNG), which measures the efficacy to suppress spatially uncorrelated noise, and directivity factor (DF), which quantifies how the microphone array performs in the environment of reverberation\cite{brandstein2013microphone}\cite{benesty2018fundamentals}.

SRP-PHAT features based on the CDDMA-beamformer are extracted to represent spatial localization of the speakers \cite{dibiase2000high}. We formalize the microphone array signals received per frame as:
\begin{equation}\label{eq:1}
\textbf{x}\left( \omega, \theta \right) = [x_{1}, x_{2},\ \cdots \ x_{M}]^{T},
\end{equation}
where the superscript $^{T}$ represents the transpose operator, $\omega = 2 \pi \mathit{f} $ is the angular frequency, $\mathit{f}$ is the temporal frequency,  $\theta$ is the incident angle of the signal, and $x_{m}$ represents the signal of each microphone. For each candidate of incident angle $\theta$, we design each corresponding CDDMA-beamformer to target at the direction of $\theta$, denoted as $\textbf{h}_{cddma}(\omega, \theta)$, and we calculate the transient steering response power(SRP) at $n^{th}$ frame as below:
\begin{equation}\label{eq:1}
\textbf{P}_{n} \left(  \theta \right) = \int_{-\infty}^{+\infty} | \textbf{x}\left( \omega, \theta \right)^{H} \textbf{h}_{cddma}(\omega, \theta)|^{2} d \omega.
\end{equation}
 The locale estimate of incident angle for current frame is given by:
\begin{equation}\label{eq:1}
\hat{\theta} = \underset{\theta}{argmax} \ \textbf{P}\left(  \theta \right)
\end{equation}
Based on the estimate of SRP at each frame, we can form a spatial spectrum  as below:
\begin{equation}\label{eq:2}
\mathcal{B} \left(  \theta , n \right) = \textbf{P}_{n} \left(  \theta \right), \forall \ n \in \mathbb{N}^+
\end{equation}

A smoothing function is applied to the estimated angle $\hat\theta$ of each frame to filter noisy direction of arrivals. 

The SRP-PHAT extracted has a dimension of 120 per frame, each representing 3 angular degree. In addition, the log energy received by each of the microphone is included. Hence the spatial feature we use for each frame has a total dimension of 128.

For acoustic feature, filterbanks are extracted from each beam formed by CDDMA.

\subsection{BeamTransformer Architecture}

\begin{figure*}[t]
  \centering
  \includegraphics[width=\linewidth]{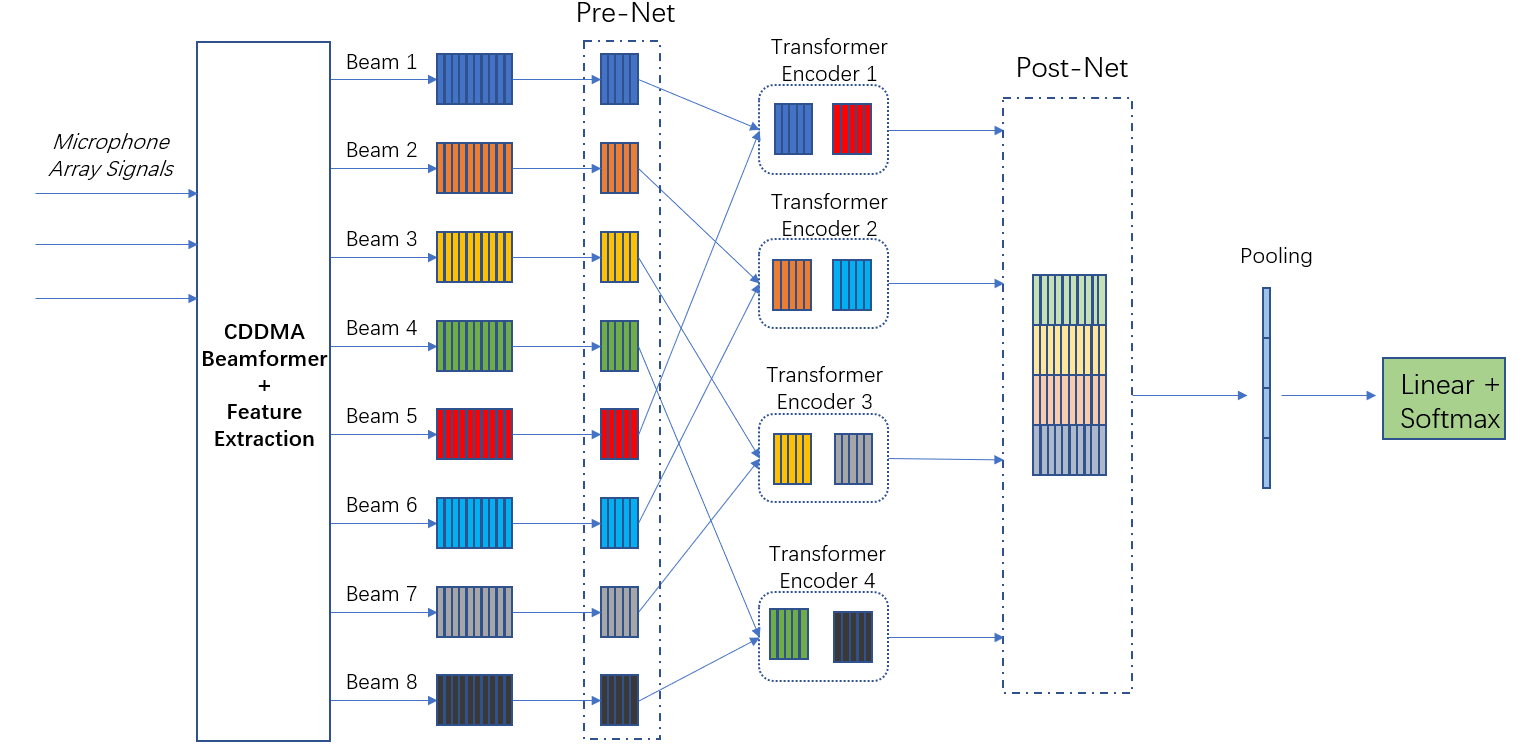}
  \caption{Architecture of BeamTransformer.}
  \label{fig:beamtransformer}
\end{figure*}

Figure \ref{fig:beamtransformer} displays the architecture of a BeamTransformer. Pointing directions from beams 1 to 8 are distributed evenly in a circular manner. Therefore, pointing direction of two adjacent beams has an angular size of 45 degree, which also indicates that beam 1 and beam 5 have opposite pointing directions.
 A 1D convolution Pre-Net is applied to beam-wise filterbank features. Pre-Net shortens the input lengths from $L$ to $\frac{L}{4}$, which reduces the computation cost of the following transformers. Beams from opposite directions are paired up. Each of the four pairs are fed to a transformer encoder. The main motivation to perform a re-combination of beams in this manner is to reduce the number of parameters and computation costs of transformer while retaining as much uncorrelated spatial information as possible to help identify sequential difference among beams. The opposite beams, such as beam 1 and beam 5, will have smallest correlations since the beamforming for speech separation have been designed to achieve relatively high front-to-back ratio. Therefore, beams on opposite direction should display more disparity than the adjacent ones, regardless of the relative localization of overlapped speakers. The outputs of transformer encoders are then stacked together and go through a Post-Net. Post-Net consists of a simple feed-forward network.

\begin{figure}[t]
\centering
\includegraphics[width=\linewidth]{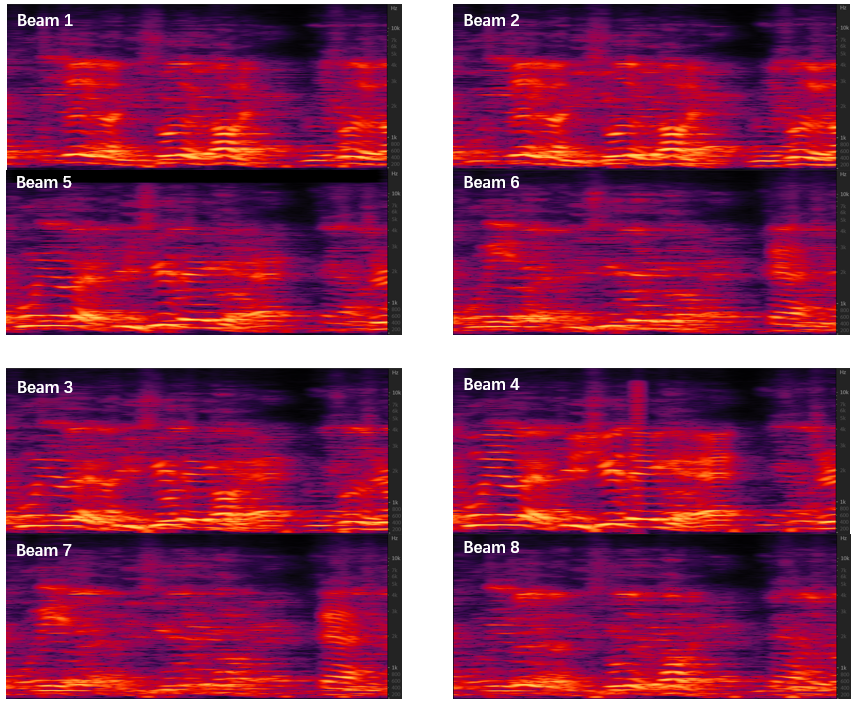}
\caption{Spectrogram of beams from overlapping segments. Two overlapping speakers are located near Beam 1 and 4, respectively.}
\label{fig:overlap}
\end{figure}

\begin{figure}[t]
\centering
\includegraphics[width=\linewidth]{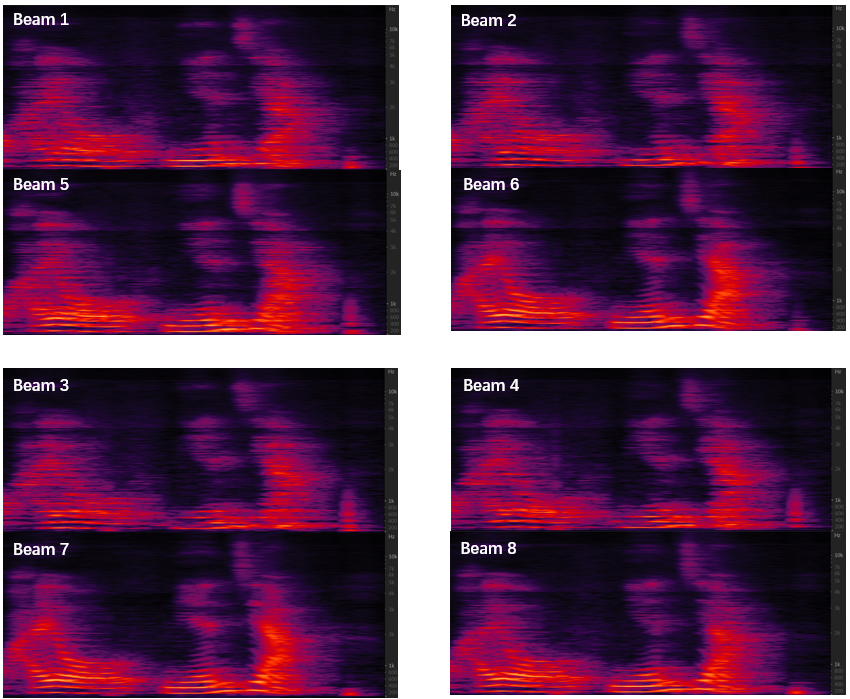}
\caption{Spectrogram of beams from non-overlapping segments. One single speaker is located near Beam 6.}
\label{fig:nonoverlap}
\end{figure}

Figure \ref{fig:overlap} and Figure \ref{fig:nonoverlap} illustrate the key feature that Beam-Transformer is trying to identify. When there are two overlapped speakers (Figure \ref{fig:overlap}), the target beams pointing to the corresponding speakers have different dominant harmonic structure of speech in the spectrogram. When there is one single speaker (Figure \ref{fig:nonoverlap}), on the other hand, there is only one dominant harmonic structure of speech, which is in the target beam pointing to the direction of speaker. As the beam direction deviates from the direction of speech, the harmonic structure gets monotonically weaker, as the signals are more suppressed by CDDMA beamforming.   

\subsection{SpatialNet Architecture}

SpatialNet is the architecture used to train the spatial spectrum input. The dimension of input is 128 each frame, including 120 dimensional SRP-PHAT feature and 8-dimensional microphone array log energy, as mentioned above. It has the same Pre-Net, Post-Net and transformer encoder structures as in BeamTransformer. The only difference is that a re-combination of beams is not required by SpatialNet.

\begin{figure}[t]
\centering
\includegraphics[width=\linewidth]{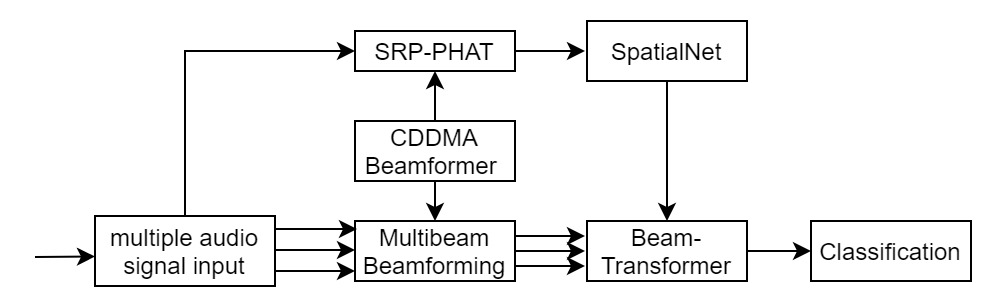}
\caption{Spatial information such as SRP-PHAT is extracted and fed into SpatialNet. The output of SpatialNet is concatenated to the Post-Net component of BeamTransformer.}
\label{fig:spatialnet}
\end{figure}

Since BeamTransformer takes the acoustic signals as input and SpatialNet takes only spatial information, they can be effective complements to each other. In Figure \ref{fig:spatialnet} we show how the SpatialNet and BeamTransformer are combined. The outputs of Post-Nets from both sides are concatenated together. A frame-alignment process is applied beforehand.

\begin{table}[h]
    \centering

        \caption{Tensor dimensions after each components of BeamTransformer and SpatialNet. F is the feature dimension. L represents the length of inputs, in terms of total number of frames. D represents the specified hidden dimension.}
\begin{tabular}{|c|c|c|}
 \hline
 
 & BeamTransformer & SpatialNet \\
 \hline
 Input & $F \times L \times$ 8 & $128 \times L \times 1$\\
 
 \hline
  Pre-Net & $D \times \frac{L}{4} \times 8$ & $D \times L \times 1$ \\
  \hline
  Recombination of beams & $(D \times \frac{L}{4} \times 2) \times 4 $ & $D \times L \times 1$\\
 \hline
  Transformer Encoders & $4D \times \frac{L}{2}$  & $D \times L$ \\
 \hline
   Post-Net & $D \times \frac{L}{2}$ & $D \times L$ \\
 \hline
    Mean Pooling & $D \times 1 $ & $D \times 1$ \\
 \hline
\end{tabular}

    \label{tab:network}
\end{table}

\section{Experiments}

\subsection{Corpus}

In this paper we experiment on both the publicly available AMI corpus and self-recorded AliMeeting corpus.

The AMI corpus data are recorded using circular array with omni-directional microphones. In order to utilize the microhpone array data of AMI corpus, the spatial spectrum are estimated based on conventional circular differential microphone array(CDMA) instead of the CDDMA. We follow the AMI Diarization Setup in \cite{DBLP:journals/corr/abs-2012-14952} to split up the train set and test set. 

The AliMeeting corpus contains 175 real meetings with human annotations. Each of the meetings has a duration of around 30 minutes. There are 4 participants in each of the meetings, 182 speakers in total. Overlapping speech consists of about 40 percent of all speech data in the corpus. The train set consists of 156 meetings and the test set contains 19 meetings. All speakers and meeting rooms in the test set are unseen in the train set. Overlapping and non-overlapping segments of at least one second are selected for both training and testing stage.

\subsection{Experiment Setup}

The baseline system is trained on original single channel audio input without beamforming. 40-dimensional filterbank features are extracted and fed into a similar network structure: a Pre-Net, one 6-layer transformer encoder, and a Post-Net. Since there is only one channel, there is no re-combination of beams involved and there is only one transformer encoder. 

We also experimented using only the spatial feature and no acoustic features. A SpatialNet is trained using 128-dimensional acoustic feature. 

BeamTransformers with 40-dimensional filterbanks and 160-dimensional high dimensional filterbanks are compared to find out how much more information BeamTransformer can mine from a more detailed sequential input.

Finally, we combine BeamTransformer and SpatialNet by stacking the outputs of Post-Nets from both networks. Since the acoustic features and spatial features are relatively independent, it is expected that such combination should see a noticeable improvement. 

\subsection{Results}

Table \ref{tab:amiresults} and Table \ref{tab:aliresults} display results from AMI corpus and AliMeeting corpus, respectively. SpatialNet represents the experiments using only the spatial feature and no acoustic features. BT(High-Res) stands for BeamTrasnformer with high resolution filterbank inputs. BT+SpatialNet represents the combination of BeamTransformer and SpatialNet.

BeamTransformer+SpatialNet is reportedly the best performed system on both AMI corpus and AliMeeting corpus, immediately followed by BeamTransformer with high resolution features. A relatively improvement of $15-16\%$ on both precision and recall is observed on both corpus. 

\begin{table}[t]
  \caption{The performance of overlapping segments detection on AMI corpus (\%).
  }
  \label{tab:amiresults}
  \centering
  
  {
  \begin{tabular}{lcccc}
  \toprule
   & Accuracy & Precision & Recall & Fscore  \\
  \midrule
  Baseline & 83.4 & 75.4 & 74.6 & 75.0 \\
  \midrule
  SpatialNet & 84.6 & 76.9 & 77.0 & 77.0 \\
  \midrule
  BeamTransformer & 88.2 & 83.1 & 81.0 & 82.0 \\
  \midrule
  BT(High-Res) & 90.0 & 85.5 & 84.5 & 85.0 \\
  \midrule
  \textbf{BT+SpatialNet} & \textbf{91.7} & \textbf{87.8} & \textbf{87.0} & \textbf{87.3} \\
  \bottomrule
\end{tabular}
}
\vspace{-2mm}
\end{table}

\begin{table}[t]
  \caption{The performance of overlapping speech segments detection on AliMeeting corpus (\%). Accuracy, Precision, Recall, and Fscore are denoted as A, P, R, Fs, respectively.
  }
  \label{tab:aliresults}
  \centering
  
  {
  \begin{tabular}{lc}
  \toprule
   & Metrics - A, P, R, Fs, lengths  \\
  \midrule
  Baseline & 79.4(A), 71.9(P), 70.7(R), 71.3(Fs), 1s \\
     & 82.9(A), 76.6(P), 75.1(R), 75.8(Fs), 2s \\
  \midrule
  SpatialNet & 81.1(A), 73.5(P), 73.6(R), 73.6(Fs), 1s \\
     & 85.1(A), 79.2(P), 79.1(R), 79.2(Fs), 2s   \\
  \midrule
  BeamTransformer & 83.6(A), 77.7(P), 76.0(R), 76.8(Fs), 1s \\
     & 87.8(A), 83.7(P), 81.9(R), 82.8(Fs), 2s   \\
  \midrule
  BT(High-Res) & 85.2(A), 80.0(P), 78.4(R), 79.2(Fs), 1s \\
     & 89.7(A), 86.1(P), 84.9(R), 85.5(Fs), 2s   \\
  \midrule
  \textbf{BT+SpatialNet} & 87.1(A), 82.5(P), 81.2(R), 81.8(Fs), 1s \\
     & \textbf{91.2(A), 88.1(P), 87.2(R), 87.6(Fs), 2s}  \\
  \bottomrule
\end{tabular}
}
\vspace{-2mm}
\end{table}

\section{Conclusions}

In this paper we investigated microphone array-based approaches in detecting overlapping speech segments. We discussed the importance of OSD in the system of speaker diarization and speech separation, and the lack of reliability in current state-of-the-art performances of frame-level OSD. We found that utilizing segment-level sequence information, along with spatial information, significantly improves the performance of overlapping speech detection. To achieve our goal, we propose a specific architecture named BeamTransformer that takes advantage of beamformer's ability in spatial filtering and transformer's renowned edge in capturing sequential knowledge.

More works could be done in the future. The reported performance was achieved by training on only 80-90 hours of data. Multi-microphone data simulation and augmentation techniques could be investigated to quickly enlarge the size of training data. Furthermore, even though the overlapping speech has been separated into different beams, a single-channel target speaker extraction approach can be applied to the corresponding beam. This way we can largely remedy the negative effects when OSD went wrong.

\bibliographystyle{IEEEtran}

\bibliography{mybib}

\begin{thebibliography}{10}
\providecommand{\url}[1]{#1}
\csname url@samestyle\endcsname
\providecommand{\newblock}{\relax}
\providecommand{\bibinfo}[2]{#2}
\providecommand{\BIBentrySTDinterwordspacing}{\spaceskip=0pt\relax}
\providecommand{\BIBentryALTinterwordstretchfactor}{4}
\providecommand{\BIBentryALTinterwordspacing}{\spaceskip=\fontdimen2\font plus
\BIBentryALTinterwordstretchfactor\fontdimen3\font minus
  \fontdimen4\font\relax}
\providecommand{\BIBforeignlanguage}[2]{{%
\expandafter\ifx\csname l@#1\endcsname\relax
\typeout{** WARNING: IEEEtran.bst: No hyphenation pattern has been}%
\typeout{** loaded for the language `#1'. Using the pattern for}%
\typeout{** the default language instead.}%
\else
\language=\csname l@#1\endcsname
\fi
#2}}
\providecommand{\BIBdecl}{\relax}
\BIBdecl

\bibitem{DBLP:journals/taslp/MiroBEFFV12}
\BIBentryALTinterwordspacing
X.~A. Mir{\'{o}}, S.~Bozonnet, N.~W.~D. Evans, C.~Fredouille, G.~Friedland, and
  O.~Vinyals, ``Speaker diarization: {A} review of recent research,''
  \emph{{IEEE} Trans. Speech Audio Process.}, vol.~20, no.~2, pp. 356--370,
  2012. [Online]. Available: \url{https://doi.org/10.1109/TASL.2011.2125954}
\BIBentrySTDinterwordspacing

\bibitem{DBLP:conf/icassp/Garcia-RomeroSS17}
\BIBentryALTinterwordspacing
D.~Garcia{-}Romero, D.~Snyder, G.~Sell, D.~Povey, and A.~McCree, ``Speaker
  diarization using deep neural network embeddings,'' in \emph{2017 {IEEE}
  International Conference on Acoustics, Speech and Signal Processing, {ICASSP}
  2017, New Orleans, LA, USA, March 5-9, 2017}.\hskip 1em plus 0.5em minus
  0.4em\relax {IEEE}, 2017, pp. 4930--4934. [Online]. Available:
  \url{https://doi.org/10.1109/ICASSP.2017.7953094}
\BIBentrySTDinterwordspacing

\bibitem{DBLP:conf/icassp/WangDWMM18}
\BIBentryALTinterwordspacing
Q.~Wang, C.~Downey, L.~Wan, P.~A. Mansfield, and I.~Lopez{-}Moreno, ``Speaker
  diarization with {LSTM},'' in \emph{2018 {IEEE} International Conference on
  Acoustics, Speech and Signal Processing, {ICASSP} 2018, Calgary, AB, Canada,
  April 15-20, 2018}.\hskip 1em plus 0.5em minus 0.4em\relax {IEEE}, 2018, pp.
  5239--5243. [Online]. Available:
  \url{https://doi.org/10.1109/ICASSP.2018.8462628}
\BIBentrySTDinterwordspacing

\bibitem{DBLP:journals/corr/abs-2012-14952}
\BIBentryALTinterwordspacing
F.~Landini, J.~Profant, M.~D{\'{\i}}ez, and L.~Burget, ``Bayesian {HMM}
  clustering of x-vector sequences (vbx) in speaker diarization: theory,
  implementation and analysis on standard tasks,'' \emph{CoRR}, vol.
  abs/2012.14952, 2020. [Online]. Available:
  \url{https://arxiv.org/abs/2012.14952}
\BIBentrySTDinterwordspacing

\bibitem{DBLP:conf/interspeech/ZhengLSL19a}
\BIBentryALTinterwordspacing
S.~Zheng, G.~Liu, H.~Suo, and Y.~Lei, ``Autoencoder-based semi-supervised
  curriculum learning for out-of-domain speaker verification,'' in
  \emph{Interspeech 2019, 20th Annual Conference of the International Speech
  Communication Association, Graz, Austria, 15-19 September 2019}, G.~Kubin and
  Z.~Kacic, Eds.\hskip 1em plus 0.5em minus 0.4em\relax {ISCA}, 2019, pp.
  4360--4364. [Online]. Available:
  \url{https://doi.org/10.21437/Interspeech.2019-1440}
\BIBentrySTDinterwordspacing

\bibitem{DBLP:conf/interspeech/ZhengLS20}
S.~Zheng, Y.~Lei, and H.~Suo, ``Phonetically-aware coupled network for short
  duration text-independent speaker verification,'' in \emph{Annual Conference
  of the International Speech Communication Association (INTERSPEECH)}, 2020,
  pp. 926--930.

\bibitem{DBLP:conf/asru/FujitaKHXNW19}
\BIBentryALTinterwordspacing
Y.~Fujita, N.~Kanda, S.~Horiguchi, Y.~Xue, K.~Nagamatsu, and S.~Watanabe,
  ``End-to-end neural speaker diarization with self-attention,'' in
  \emph{{IEEE} Automatic Speech Recognition and Understanding Workshop, {ASRU}
  2019, Singapore, December 14-18, 2019}.\hskip 1em plus 0.5em minus
  0.4em\relax {IEEE}, 2019, pp. 296--303. [Online]. Available:
  \url{https://doi.org/10.1109/ASRU46091.2019.9003959}
\BIBentrySTDinterwordspacing

\bibitem{DBLP:conf/interspeech/FujitaKHNW19}
\BIBentryALTinterwordspacing
Y.~Fujita, N.~Kanda, S.~Horiguchi, K.~Nagamatsu, and S.~Watanabe, ``End-to-end
  neural speaker diarization with permutation-free objectives,'' in
  \emph{Interspeech 2019, 20th Annual Conference of the International Speech
  Communication Association, Graz, Austria, 15-19 September 2019}, G.~Kubin and
  Z.~Kacic, Eds.\hskip 1em plus 0.5em minus 0.4em\relax {ISCA}, 2019, pp.
  4300--4304. [Online]. Available:
  \url{https://doi.org/10.21437/Interspeech.2019-2899}
\BIBentrySTDinterwordspacing

\bibitem{DBLP:conf/interspeech/HoriguchiF0XN20}
\BIBentryALTinterwordspacing
S.~Horiguchi, Y.~Fujita, S.~Watanabe, Y.~Xue, and K.~Nagamatsu, ``End-to-end
  speaker diarization for an unknown number of speakers with encoder-decoder
  based attractors,'' in \emph{Interspeech 2020, 21st Annual Conference of the
  International Speech Communication Association, Virtual Event, Shanghai,
  China, 25-29 October 2020}, H.~Meng, B.~Xu, and T.~F. Zheng, Eds.\hskip 1em
  plus 0.5em minus 0.4em\relax {ISCA}, 2020, pp. 269--273. [Online]. Available:
  \url{https://doi.org/10.21437/Interspeech.2020-1022}
\BIBentrySTDinterwordspacing

\bibitem{DBLP:journals/corr/abs-2003-02966}
\BIBentryALTinterwordspacing
Y.~Fujita, S.~Watanabe, S.~Horiguchi, Y.~Xue, and K.~Nagamatsu, ``End-to-end
  neural diarization: Reformulating speaker diarization as simple multi-label
  classification,'' \emph{CoRR}, vol. abs/2003.02966, 2020. [Online].
  Available: \url{https://arxiv.org/abs/2003.02966}
\BIBentrySTDinterwordspacing

\bibitem{DBLP:conf/interspeech/WangLSWCLHLPNG20}
Q.~Wang, I.~Lopez{-}Moreno, M.~Saglam, K.~W. Wilson, A.~Chiao, R.~Liu, Y.~He,
  W.~Li, J.~Pelecanos, M.~Nika, and A.~Gruenstein, ``Voicefilter-lite:
  Streaming targeted voice separation for on-device speech recognition,'' in
  \emph{Interspeech 2020, 21st Annual Conference of the International Speech
  Communication Association, Virtual Event, Shanghai, China, 25-29 October
  2020}, H.~Meng, B.~Xu, and T.~F. Zheng, Eds., pp. 2677--2681.

\bibitem{DBLP:conf/icassp/DelcroixZKON18}
M.~Delcroix, K.~Zmol{\'{\i}}kov{\'{a}}, K.~Kinoshita, A.~Ogawa, and
  T.~Nakatani, ``Single channel target speaker extraction and recognition with
  speaker beam,'' in \emph{{IEEE} International Conference on Acoustics, Speech
  and Signal Processing, {ICASSP} 2018, Calgary, AB, Canada, April 15-20,
  2018}.\hskip 1em plus 0.5em minus 0.4em\relax {IEEE}, 2018, pp. 5554--5558.

\bibitem{DBLP:journals/taslp/LuoM19}
\BIBentryALTinterwordspacing
Y.~Luo and N.~Mesgarani, ``Conv-tasnet: Surpassing ideal time-frequency
  magnitude masking for speech separation,'' \emph{{IEEE} {ACM} Trans. Audio
  Speech Lang. Process.}, vol.~27, no.~8, pp. 1256--1266, 2019. [Online].
  Available: \url{https://doi.org/10.1109/TASLP.2019.2915167}
\BIBentrySTDinterwordspacing

\bibitem{DBLP:journals/taslp/WangNW14}
\BIBentryALTinterwordspacing
Y.~Wang, A.~Narayanan, and D.~Wang, ``On training targets for supervised speech
  separation,'' \emph{{IEEE} {ACM} Trans. Audio Speech Lang. Process.},
  vol.~22, no.~12, pp. 1849--1858, 2014. [Online]. Available:
  \url{https://doi.org/10.1109/TASLP.2014.2352935}
\BIBentrySTDinterwordspacing

\bibitem{DBLP:journals/taslp/WangC18a}
\BIBentryALTinterwordspacing
D.~Wang and J.~Chen, ``Supervised speech separation based on deep learning: An
  overview,'' \emph{{IEEE} {ACM} Trans. Audio Speech Lang. Process.}, vol.~26,
  no.~10, pp. 1702--1726, 2018. [Online]. Available:
  \url{https://doi.org/10.1109/TASLP.2018.2842159}
\BIBentrySTDinterwordspacing

\bibitem{DBLP:conf/interspeech/WangMWSWHSWJL19}
\BIBentryALTinterwordspacing
Q.~Wang, H.~Muckenhirn, K.~W. Wilson, P.~Sridhar, Z.~Wu, J.~R. Hershey, R.~A.
  Saurous, R.~J. Weiss, Y.~Jia, and I.~Lopez{-}Moreno, ``Voicefilter: Targeted
  voice separation by speaker-conditioned spectrogram masking,'' in
  \emph{Interspeech 2019, 20th Annual Conference of the International Speech
  Communication Association, Graz, Austria, 15-19 September 2019}, G.~Kubin and
  Z.~Kacic, Eds.\hskip 1em plus 0.5em minus 0.4em\relax {ISCA}, 2019, pp.
  2728--2732. [Online]. Available:
  \url{https://doi.org/10.21437/Interspeech.2019-1101}
\BIBentrySTDinterwordspacing

\bibitem{anguera2007acoustic}
X.~Anguera, C.~Wooters, and J.~Hernando, ``Acoustic beamforming for speaker
  diarization of meetings,'' \emph{IEEE Transactions on Audio, Speech, and
  Language Processing}, vol.~15, no.~7, pp. 2011--2022, 2007.

\bibitem{DBLP:conf/interspeech/ZhengS21}
S.~Zheng, W.~Huang, X.~Wang, H.~Suo, J.~Feng, and Z.~Yan, ``A real-time speaker
  diarization system based on spatial spectrum,'' in \emph{2021 {IEEE}
  International Conference on Acoustics, Speech and Signal Processing, {ICASSP}
  2021, Toronto, Canada, June 6-11}, 2021.

\bibitem{DBLP:conf/interspeech/YoshiokaECXA18}
\BIBentryALTinterwordspacing
T.~Yoshioka, H.~Erdogan, Z.~Chen, X.~Xiao, and F.~Alleva, ``Recognizing
  overlapped speech in meetings: {A} multichannel separation approach using
  neural networks,'' in \emph{Interspeech 2018, 19th Annual Conference of the
  International Speech Communication Association, Hyderabad, India, 2-6
  September 2018}, B.~Yegnanarayana, Ed.\hskip 1em plus 0.5em minus 0.4em\relax
  {ISCA}, 2018, pp. 3038--3042. [Online]. Available:
  \url{https://doi.org/10.21437/Interspeech.2018-2284}
\BIBentrySTDinterwordspacing

\bibitem{mccowan2005ami}
I.~McCowan, J.~Carletta, W.~Kraaij, S.~Ashby, S.~Bourban, M.~Flynn,
  M.~Guillemot, T.~Hain, J.~Kadlec, V.~Karaiskos \emph{et~al.}, ``The ami
  meeting corpus,'' in \emph{Proceedings of the 5th International Conference on
  Methods and Techniques in Behavioral Research}, vol.~88.\hskip 1em plus 0.5em
  minus 0.4em\relax Citeseer, 2005, p. 100.

\bibitem{DBLP:conf/interspeech/AndreiCB17}
\BIBentryALTinterwordspacing
V.~Andrei, H.~Cucu, and C.~Burileanu, ``Detecting overlapped speech on short
  timeframes using deep learning,'' in \emph{Interspeech 2017, 18th Annual
  Conference of the International Speech Communication Association, Stockholm,
  Sweden, August 20-24, 2017}, F.~Lacerda, Ed.\hskip 1em plus 0.5em minus
  0.4em\relax {ISCA}, 2017, pp. 1198--1202. [Online]. Available:
  \url{http://www.isca-speech.org/archive/Interspeech\_2017/abstracts/0188.html}
\BIBentrySTDinterwordspacing

\bibitem{DBLP:conf/specom/KunesovaHZR19}
M.~Kunesov{\'{a}}, M.~Hr{\'{u}}z, Z.~Zaj{\'{\i}}c, and V.~Radov{\'{a}},
  ``Detection of overlapping speech for the purposes of speaker diarization,''
  in \emph{Speech and Computer - 21st International Conference, {SPECOM} 2019,
  Istanbul, Turkey, August 20-25, 2019, Proceedings}, ser. Lecture Notes in
  Computer Science, A.~A. Salah, A.~Karpov, and R.~Potapova, Eds., vol.
  11658.\hskip 1em plus 0.5em minus 0.4em\relax Springer, 2019, pp. 247--257.

\bibitem{DBLP:conf/icassp/BullockBG20}
\BIBentryALTinterwordspacing
L.~Bullock, H.~Bredin, and L.~P. Garc{\'{\i}}a{-}Perera, ``Overlap-aware
  diarization: Resegmentation using neural end-to-end overlapped speech
  detection,'' in \emph{2020 {IEEE} International Conference on Acoustics,
  Speech and Signal Processing, {ICASSP} 2020, Barcelona, Spain, May 4-8,
  2020}.\hskip 1em plus 0.5em minus 0.4em\relax {IEEE}, 2020, pp. 7114--7118.
  [Online]. Available: \url{https://doi.org/10.1109/ICASSP40776.2020.9053096}
\BIBentrySTDinterwordspacing

\bibitem{huangdifferential}
W.~Huang and J.~Feng, ``Differential beamforming for uniform circular array
  with directional microphones,'' in \emph{INTERSPEECH 2020}.

\bibitem{brandstein2013microphone}
M.~Brandstein and D.~Ward, \emph{Microphone arrays: signal processing
  techniques and applications}.\hskip 1em plus 0.5em minus 0.4em\relax Springer
  Science \& Business Media, 2013.

\bibitem{benesty2018fundamentals}
J.~Benesty, I.~Cohen, and J.~Chen, \emph{Fundamentals of Signal Enhancement and
  Array Signal Processing}.\hskip 1em plus 0.5em minus 0.4em\relax Wiley Online
  Library, 2018.

\bibitem{dibiase2000high}
J.~H. DiBiase, \emph{A high-accuracy, low-latency technique for talker
  localization in reverberant environments using microphone arrays}.\hskip 1em
  plus 0.5em minus 0.4em\relax Brown University Providence, RI, 2000.

\end{thebibliography}


\end{document}